**Structural, electronic, magnetic, and thermal properties of single-crystalline UNi$_{0.5}$Sb$_2$**

M. S. Torikachvili,[1] B. K. Davis,[1,2] K. Kothapalli,[3] H. Nakotte,[3,4] A. J. Schultz,[5] E. D. Mun,[6] and S. L. Bud'ko[6]

[1] Department of Physics, San Diego State University, San Diego, CA 92182-1233
[2] Quantum Design, 6325 Lusk Boulevard, San Diego, CA 92121
[3] Department of Physics, New Mexico State University, Las Cruces, NM 88003
[4] LANSCE, Los Alamos National Laboratory, Los Alamos, NM 87545
[5] X-ray Science Division, Argonne National Laboratory, Argonne, IL 60439
[6] Ames Laboratory and Department of Physics and Astronomy, Iowa State University, Ames, IA 50011

*Abstract*

We studied the properties of the antiferromagnetic (AFM) UNi$_{0.5}$Sb$_2$ ($T_N \approx 161$ K) compound in Sb flux-grown single crystals by means of measurements of neutron diffraction, magnetic susceptibility ($\chi$), specific heat ($C_p$), thermopower ($S$), thermal conductivity ($\kappa$), linear thermal expansion ($\Delta L/L$), and electrical resistivity ($\rho$) under hydrostatic pressures ($P$) up to 22 kbar. The neutron diffraction measurements revealed that the compound crystallizes in the tetragonal $P4_2/nmc$ structure, and the value of the U-moments yielded by the data at 25 K is $\approx 1.85 \pm 0.12$ $\mu_B$/U-ion. In addition to the features in the bulk properties observed at $T_N$, two other hysteretic features centered near 40 and 85 K were observed in the measurements of $\chi$, $S$, $\rho$, and $\Delta L/L$. Hydrostatic pressure was found to raise $T_N$ at the rate of $\approx 0.76$ K/kbar, while suppressing the two low temperature features. These features are discussed in the context of Fermi surface and hybridization effects.

## 1- Introduction

A large number of uranium-based intermetallic ternary compounds with general composition U$T$Sb$_2$ ($T$ = transition metal) have been reported to crystallize in the tetragonal HfCuSi$_2$-type structure (space group $P4/nmm$, No. 129).[1] In this structure, planar layers of Sb, $T$, and U-Sb are stacked along the c-axis, conferring these materials with strongly anisotropic properties. The U$T$Sb$_2$ compounds have been reported to order ferromagnetically (FM) for $T$ = Co, Cu, Ag, and Au, and antiferromagnetically (AFM) for $T$ = Ni, Ru, and Pd, in temperatures ranging from ≈ 35 to 200 K.[1,2] It is important to point out that many $RT$Sb$_2$ ($R$ = rare earth) compounds form with the same crystal structure, and they also show rich and complex magnetic behavior.[3]

The focus of this paper is on UNi$_{0.5}$Sb$_2$, a particularly noteworthy case, as will be outlined below. Previous work by Kaczorowski et al. [Ref. 1] on polycrystalline specimens, with nominal composition UNiSb$_2$, showed the occurrence of AFM order with $T_N$ ≈ 175 K. The magnetic structure was described as being formed by U-moments coupled antiferromagnetically along the $c$-axis, with the moments in each $ab$-plane having the same orientation. The effective moment inferred from measurements of the temperature ($T$) dependence of the magnetic susceptibility was 2.46 $\mu_B$, which is somewhat reduced from the expected values of 3.58 and 3.62 $\mu_B$, for U$^{4+}$ and U$^{3+}$, respectively. Since the shortest U-U distance is ≈ 4.3 Å, which is well above the Hill limit of ≈ 3.6 Å,[4] the amount of $f$-$f$ hybridization is expected to be negligible, and the magnetic properties are determined by the $f$-$s$, $f$-$p$ and $f$-$d$ hybridization of the U-ions with Ni and Sb.

Following the studies in polycrystalline specimens, attempts of growing single-crystals from Sb flux were successfully carried out by three different groups.[5-7] However, the detailed crystallographic analysis by Bukowski et al.,[5] using X-ray diffraction (XRD) data, found that the growth from a ≈ 1:1:14 ratio of U:Ni:Sb, aimed at obtaining crystals with 1:1:2 composition, yielded crystals with the same structure as the polycrystalline samples of Ref. 1, but with half occupancy at the Ni-site, i.e. the actual composition was UNi$_{0.5}$Sb$_2$.[5] This composition was confirmed by Davis et al., using XRD data from a single crystal diffractometer.[7] Although the nominal composition of the Sb-grown crystals by Ikeda et al. is UNiSb$_2$,[6] no crystallographic information is provided. In light of tendency for half occupancy of the Ni site in Sb flux-grown crystals,[5,7] and the similarity of the electrical resistivity and magnetic susceptibility data in Refs. 6 and 7, it is quite probable that the actual composition is close to $UNi_{0.5}Sb_2$. Partial occupancy at the $T$ sites in single crystals has been reported also for $UCo_{0.5}Sb_2$,[2] $UCu_{0.44}Sb_2$,[8] and $UPd_{0.6}Sb_2$.[9]

Bulk measurements of electrical resistivity ($\rho$), magnetic susceptibility ($\chi$), specific heat ($C_p$), thermal conductivity ($\kappa$), and thermoelectric power ($S$) in Sb flux-grown $UNi_{0.5}Sb_2$ single crystals revealed features that are consistent with the onset of AFM near $\approx$ 161 K.[5-7, 10-12] For example, the $\rho(T)$ data with current in the $ab$-plane show a sharp drop near $T_N$, consistent with the loss of spin-disorder scattering across the $ab$-plane in the ordered phase.[5-7] However, in addition to the features in all bulk properties near $T_N$, one (Ref. 5) or two (Refs. 6-7) much smaller features were observed at lower temperatures. Measurements by Bukowski and collaborators showed a small feature centered near 60 K in $\rho(T)$,[5] $\chi(T)$,[5] $\kappa(T)$,[10] $C_p(T)$,[11,12] and $S(T)$.[12] This feature was clearly hysteretic upon the cooling/warming cycles.[12] On the other hand, measurements $\rho(T)$ and $\chi(T)$ by Ikeda et al.,[6] and Davis et al.,[7] in samples from their own groups, instead of the feature at 60 K, showed 2 small hysteretic features centered near 40 K and 85 K. Contrary to the 60 K feature in Refs. [11,12], these two low-$T$ features didn't leave any obvious signatures in the $C_p(T)$ data of Ref. 7. These two features were found to be suppressed in the $\rho_{ab}(T)$ data under hydrostatic pressures of $\approx$ 20 kbar.[7]

In spite of the seemingly high quality of the crystals from the three groups, and the similar behaviors near $T_N$, there are some discrepancies in regard to the onset of the hysteretic features at low-$T$. These features are not reminiscent of observations in any other $UTSb_2$ or $RTSb_2$ compounds, and their origin is still elusive. However, the effect of pressure is to suppress the instability that drives them, and to stabilize the high temperature phase.[7] In light of the Ni deficiency of this structure, it is fair to question whether the low-$T$ instabilities are intrinsic to the $UNi_{0.5}Sb_2$ phase, or perhaps due to extrinsic effects, as for example, small variations in Ni occupancy throughout the sample.

In order to further our understanding of the structural and magnetic properties of $UNi_{0.5}Sb_2$, and to probe the underlying mechanisms associated with the features in the bulk properties below $T_N$ in particular, we carried out an in-depth study consisting of the determination of the crystalline and magnetic structures from neutron single crystal diffraction data, and measurements of $\rho_{ab}(T)$ and $\rho_c(T)$ under hydrostatic pressures up to $\approx$ 22 kbar, magnetic susceptibility, thermal expansion, specific heat, thermal conductivity, and thermopower. These studies were all performed on samples from the same batch.

**2- Experimental details**

The UNi$_{0.5}$Sb$_2$ single crystals for this work were grown from a Sb-rich flux, using a technique described in Ref. 13. The crystals have a plate-like appearance with the *c*-axis perpendicular to the plate, and typical dimensions of 2 x 2 x 0.5 mm$^3$. Powder XRD analysis on crushed crystals showed that they are single-phase. Neutron diffraction data collected on the Single Crystal Diffractometer (SCD)[14] at the Los Alamos Neutron Science Center (LANSCE) permitted a more detailed refinement of the crystal structure at ambient temperature, as it will be discussed in the next section. Using neutron time-of-flight, the SCD can be viewed as a wavelength-resolved Laue camera, where the pulsed beam of neutrons generated by the spallation source and scattered by the crystal are captured by two $^6$Li area detectors that cover a substantial fraction of the reciprocal space at one time. Therefore the SCD is well-suited for the study of unknown crystal and magnetic structures. SCD data were collected also at 25 K in order verify the crystal structure, and resolve the configuration of the U-moments in the AFM state. The measurement of all electronic, magnetic, and thermal properties were preformed using a Quantum Design Physical Property Measurements System (PPMS), which served as a variable temperature and magnetic field platform. The measurements of $\rho$ (*T*, *P*) in hydrostatic pressures up to 22 kbar were carried out using a Be-Cu self-clamping piston-cylinder pressure cell, with a hardened NiCrAl-alloy core. Four Pt leads were attached to the sample using Epotek H20E Ag-loaded epoxy. For the $\rho_c$ measurements, both the voltage and current leads were attached to same facet of the crystal, which caused some uncertainty in the determination of the absolute value of the resistivity. The sample leads, and coils of manganin and Pb, which served as high-*T* and low-*T* manometers, respectively, were soldered to the tips of 12 Cu wires at the end of a Stycast-sealed feedthrough. This assembly was inserted in a Teflon cup filled with a 40:60 mixture of mineral oil:n-pentane, which served as the pressure-transmitting medium. Pressure was applied and locked in at ambient temperature with a hydraulic press, using the manganin manometer as a reference. The actual pressure in this type of cell is known to decrease with temperature, before stabilizing below ≈ 90 K,[15] due to the different thermal expansion properties of the cell's constituents. The pressure at low temperatures was determined from the superconducting transition temperature of the Pb manometer, whereas the pressure values between ambient temperature and 90 K were estimated by linear interpolation. The pressure cell was fit to the PPMS, and the temperature of the sample was inferred from a calibrated Cernox sensor (CX-1030-SD) attached to the body of the cell. In order to minimize the temperature lag between sample and sensor, the temperature sweep rate was kept below 0.25 K/min, which yielded negligible differences in sample temperature between the cooling and warming cycles. The magnetization and specific heat measurements were performed with the PPMS vibrating sample magnetometer, and relaxation calorimeter, respectively. For the thermopower and thermal conductivity measurements, which were carried out with the thermal transport option of the PPMS, a sample with approximate dimensions of 1.5 x 1.5 x 0.15 mm$^3$ was laid across two contact strips, and a temperature gradient of ≈ 2% of the sample temperature was established across the *ab*-plane, between the hot and cold ends of the sample. Thermal expansion measurements were performed with a capacitive dilatometer built with oxygen-free-high-conductivity (OFHC) copper.[16] In light of the strong thermal hysteresis of the low temperature features, most measurements were performed both upon cooling and warming.

**3- Experimental results**

**3.1- Structural determination by neutron diffraction**

The geometrical configuration of SCD allows three-dimensional samplings of diffracted intensities over a hemisphere in reciprocal space, using just a few angular settings of the $\chi$ and $\phi$ angles with an automated Huber goniometer. Ambient temperature data on a UNi$_{0.5}$Sb$_2$ single crystal with approximate dimensions 2 x 2 x 0.5 mm$^3$ were collected for 11 angular settings (histograms), 4 hours each. Statistically significant intensities ($\geq 3\sigma$, where $\sigma$ is the standard deviation of the background) were found for 243 unique (*hkl*) reflections with limiting indices $-5 \leq h \leq 5$, $-5 \leq k \leq 5$ and $-25 \leq l \leq 24$. The integrated intensities of the diffraction spots were computed using subroutines from the Integrated Spectral Analysis Workbench (ISAW) software.[17] The structural parameters could then be determined by fitting the observed integrated intensities to the structural models using the Generalized Structural Analysis Software (GSAS).[18]

Although previously reported structural studies by means of X-ray diffraction, in Sb flux-grown UNi$_{0.5}$Sb$_2$ single crystals, suggested a tetragonal HfCuSi$_2$-type structure (space group: *P*4/*nmm*; No. 129), with lattice parameters $a \approx 4.319$ Å and $c \approx 9.015$ Å, and only half occupancy at the Ni site,[5] these studies revealed a few additional weak ($h$, $k$, $l/2$)-type reflections, assuming the HfCuSi$_2$-type structure, which were not included in the

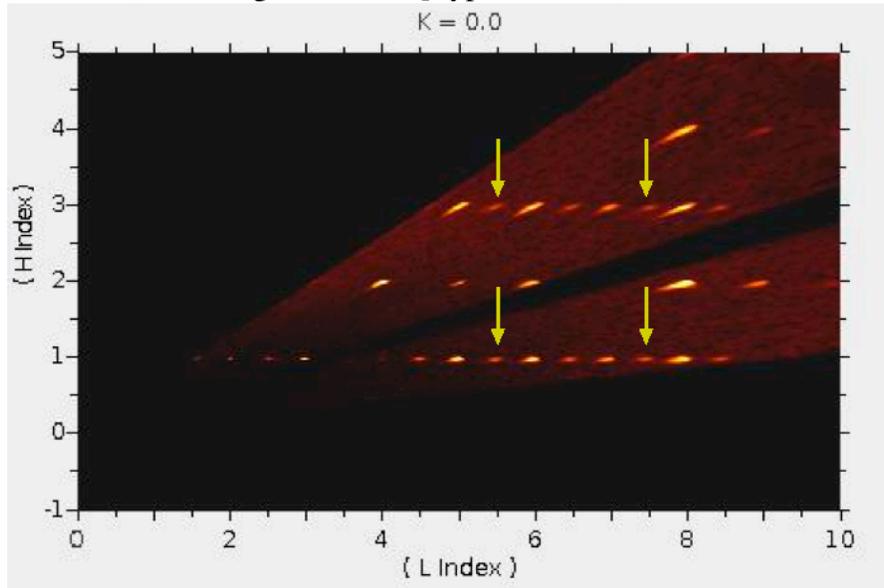

Figure 1 (color online) – A portion of the ($h0l$) plane for UNi$_{0.5}$Sb$_2$ at ambient temperature, indexed according to the HfCuSi$_2$-type structure. The arrows indicate 4 of the weaker reflections with $l = (2n+1)/2$ indices. The $l$ indices are integers when indexed according to the $P4_2/nmc$ structure. The dark band across the pane is due to the space between the two detectors.

structural refinement.[5] Albeit carried out on a different single-crystal, the present neutron diffraction identified a few weaker ($h$, $k$, $l/2$)-type reflections that could not be accounted for with the HfCuSi$_2$ model as well. Since $\lambda/2$ contamination is not an issue in time-of-flight studies, these weaker reflections should be regarded as intrinsic, and other structural models should be considered. Least squares structural refinements using GSAS (Ref. 18) and SHELX (Ref. 19), taking into account the reflections of all intensities, i.e. including the outliers of the HfCuSi$_2$ model, yielded best fits assuming a tetragonal structure represented by the space group $P4_2/nmc$ (No. 137), with room temperature (RT) lattice parameters $a \approx 4.333(6)$ Å and $c \approx 17.868(4)$ Å. Unlike $P4/nmm$, the $P4_2/nmc$ structure indexes all the observed reflections. An example of a portion of the ($h0l$) plane in reciprocal space from one of the room temperature histograms is shown in Fig. 1. The refinement of the atomic positions for RT, assuming the $P4_2/nmc$ structure, yielded the positional parameters listed in Table 1. It should be noted that there are two distinct Ni positions in $P4_2/nmc$, 2a and 2b, and that the fits indicate that they are unoccupied and fully occupied, respectively. A diagram of the refined structure with the Ni completely occupying only the 2b position is shown in Fig. 2. The unoccupied 2a sites are indicated by the symbols in a faded color. A marginally improved fit for the RT data was obtained when the occupancy of the 2a site was unconstrained from its zero value, yielding 0.08 and 0.92 as the occupancies of the 2a and 2b sites, respectively. However, this small improvement ($R_w$ was reduced from $\approx$ 16.8 to $\approx$ 16.2%) when additional fitting parameters are included in the least-squares fits of GSAS maybe misleading, and it should be considered cautiously.

Table 1 - Structural parameters for $UNi_{0.5}Sb_2$ at ambient temperature. The U and Sb occupancies were fixed to 1.

| Space group: $P4_2/nmc$ (No. 137) - Tetragonal | | | | | |
|---|---|---|---|---|---|
| Lattice parameters: $a = 4.333(6)$ Å ; $c = 17.868(4)$ Å | | | | | |
| $\chi^2 = 1.365$, $R(F^2) = 7.77\%$, $R_w(F^2) = 16.83\%$ | | | | | |
| *atom* | *position* | *x* | *y* | *z* | *occupancy* |
| U | 4d | 1/4 | 1/4 | 0.1338(7) | 1.0 |
| Sb1 | 4c | 3/4 | 1/4 | 0 | 1.0 |
| Sb2 | 4d | 3/4 | 3/4 | 0.1773(1) | 1.0 |
| Ni1 | 2b | 3/4 | 1/4 | 1/4 | 1.0 |
| Ni2 | 2a | 1/4 | 3/4 | 1/4 | 0.0 |

Although the goodness-of-fitness value obtained for the $P4_2/nmc$ model in this work is about 50% higher than for $P4/nmm$, the relatively good fit of the $UNi_{0.5}Sb_2$ structure to the latter, in spite of its inability to account for the weaker 1/2-indexed reflections, suggests some similarities between these models. A careful comparison of the

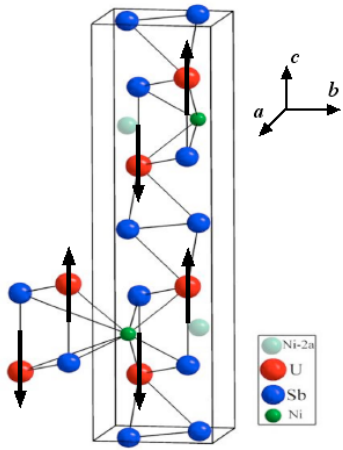

Figure 2 (color online) – Schematic diagram of the crystal and magnetic structures for $UNi_{0.5}Sb_2$. The box represents the unit cell, while the few additional atoms outside of the unit cell are drawn for clarity. The faded green spheres represent the unoccupied 2a Ni positions in space group $P4_2/nmc$ (see Table 1). The arrows indicate the unit-cell configuration of the U-moments at 25 K.

two structures shows that the $P4_2/nmc$ can be obtained from the $P4/nmm$ cell by doubling along $c$, and translating the origin of the unit-cell.

**3.2- Low temperature magnetic structure**

In order to determine the magnetic structure of the AFM phase, we carried out neutron diffraction measurements in SCD at 25 K, which is well below $T_N$. The histograms at 25 K revealed additional reflections, which are clearly due to the onset of magnetic order because they occur only at larger $d$-spacings, consistent with the scattering vector $q$-dependence of the magnetic form factor. It is important to point out that both, the observed purely magnetic reflection, as well as additional intensities to the nuclear peaks, could be labeled with integer Miller indices for the $P4_2/nmc$ structure, which is evidence that the magnetic propagation vector is $Q = (0, 0, 0)$, i.e. the magnetic order is commensurate with the lattice.

All planes yielding magnetic contributions to the reflections at higher $d$-spacings can be indexed as $(h, 0, 0)$, $(0, k, 0)$, and $(h, k, 0)$, and no magnetic contributions to $(0, 0, l)$-type reflections could be identified. This is clear evidence that the magnetic moments are aligned along the $c$-axis. After corrections were made to account for variations in the incident beam intensity as a function of the wavelength, detector efficiency, absorption, thermal parameters, and extinction, we can extract the square of the magnetic structure factors from the measured intensities. Using an experimental $U^{3+}$ magnetic form factor determined for UPdSn,[20] the parallel refinement of the nuclear and magnetic structures using GSAS yielded $1.85 \pm 0.12$ $\mu_B$ for the value of U magnetic moment (at 25 K) with alternating antiferromagnetic coupling between the moments of neighboring $ab$-planes, as shown in Fig. 2. Within the experimental resolution of SCD, we do not find evidence for a structural transition for $UNi_{0.5}Sb_2$ at low temperatures, i.e. the crystal structure remains $P4_2/nmc$ down to 25 K.

The extracted structure factors for some purely magnetic and mixed reflections and their computed values are listed in Table 2, assuming the moment configuration shown in Fig. 2. The computed values show a reasonable overall agreement with the experimental values. The magnetic structure of Fig. 2 differs slightly from the structure proposed in Ref. 1. While both models consist of U-moments aligned antiferromagnetically along the $c$-direction, and coupled ferromagnetically across the tetragonal basal plane, the stacking order in Ref. 1 is "+--+", while the stacking order in Fig. 2 is "+-+-".

**Table 2** - Partial list of magnetic reflections at 25 K for UNi$_{0.5}$Sb$_2$. The Miller indices, $d$-spacings, $q$-vector magnitudes, square of the observed structure factors ($F_o$), their experimental error ($\sigma$), and the square of the computed structure factors ($F_c$) are given. The purely magnetic reflections are indicated with an asterisk. The nuclear contribution to the $F_c$ values of the mixed reflections is given in parenthesis.

| h  | k  | l  | d-spacing (Å$^{-1}$) | q (Å$^{-1}$) | $F_o^2$ | $\sigma(F_o)$ | $F_c^2$ |
|----|----|----|----------------------|--------------|---------|---------------|---------|
| *1 | 0  | 0  | 4.2980               | 0.0185       | 555     | 24            | 539     |
| *3 | 0  | 0  | 1.4320               | 0.0555       | 462     | 9             | 436     |
| *5 | 0  | 0  | 0.8596               | 0.0926       | 268     | 16            | 265     |
| *1 | 5  | -9 | 0.7771               | 0.1024       | 60      | 7             | 78      |
| *0 | 6  | -3 | 0.7111               | 0.1119       | 67      | 8             | 101     |
| 1  | -2 | -8 | 1.4585               | 0.0544       | 273     | 10            | 392(82) |
| 0  | 1  | -8 | 1.9947               | 0.0398       | 307     | 9             | 372(119)|
| -3 | 2  | -8 | 1.0497               | 0.0755       | 231     | 19            | 239(151)|

### 3.3- Magnetic susceptibility

The behavior of the magnetic susceptibility $\chi$ vs $T$ and $\chi^{-1}$ vs $T$, with $\chi$ values taken from the $M/H$ ratios, is shown in Fig. 3. These data show clearly the onset of AFM order near 161 K. Additionally, two small features centered near 40 and 85 K can be observed; these features are hysteretic in regard to the cooling and warming cycles, with the higher values of $\chi$ observed upon cooling. A fit of the $\chi^{-1}$ vs $T$ data for $T > 250$ K to a Curie-Weiss expression $\chi = \chi_0 + C/(T-\Theta)$ yielded the effective moments of $\mu_{eff} \approx 3.15\ \mu_B$ ($H//c$) and 3.25 $\mu_B$ ($H\perp c$), somewhat reduced from the U$^{4+}$ and U$^{3+}$ values of 3.58 $\mu_B$ and 3.62 $\mu_B$, respectively. The values of $\Theta$ are $\approx -240$ K ($H//c$) and $\approx 100$ K ($H\perp c$), consistent with the AFM arrangement of Fig. 2. Magnetization curves for H//c show linear behavior up to 9 T as shown in the inset of Fig. 3b.

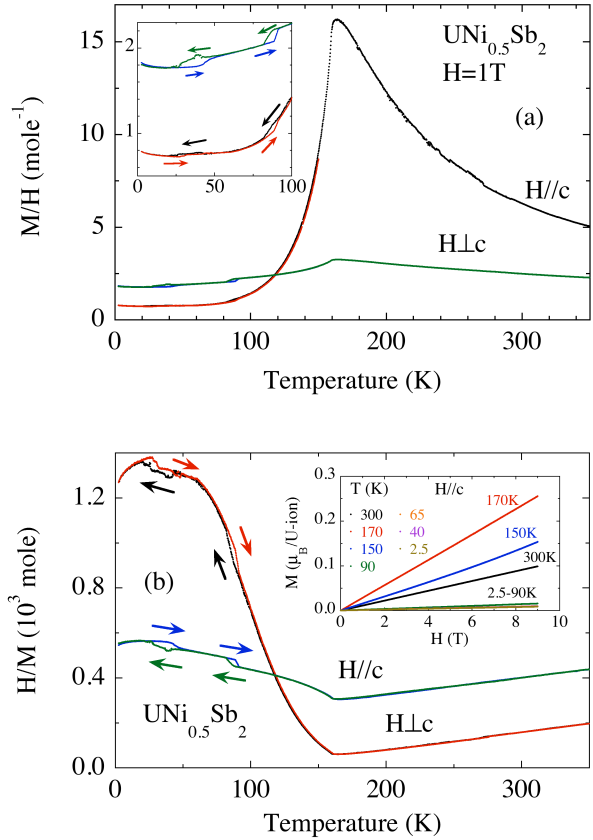

Figure 3 (color online) – (a) $\chi(T)$; and (b) $\chi^{-1}(T)$ for $UNi_{0.5}Sb_2$ for $H = 1$ T parallel and perpendicular to the *c*-axis. These data yield $T_N \approx 161$ K. Fits of $\chi(T)$ to a Curie-Weiss law yield $\Theta \approx -240$ K and $\mu_{eff} \approx 3.15$ $\mu_B$ for $H//c$, and $\Theta \approx 100$ K and $\mu_{eff} \approx 3.25$ $\mu_B$ for $H \perp c$.

### 3.4- Thermal properties

The $C_p(T)$ data for $UNi_{0.5}Sb_2$ shown in Fig. 4 display a pronounced peak near $T_N \approx 161$ K, consistent with the onset of AFM order. Missing $C_p(T)$ data on a non-magnetic analog, a rough estimate of the amount of entropy loss $\Delta S_{mag}$ due to the formation of the AFM state can be calculated by subtracting the estimated magnetic contribution between 100 and 170 K from the total entropy, as shown in the lower inset of Fig. 4. This calculation yields $\Delta S_{mag} \approx 3.0$ J/mole K, corresponding to $0.52R\ln 2$, which suggests that the magnetic order is long-range. An extrapolation of the $C/T$ vs $T^2$ data at low temperatures to $T = 0$ yields a Sommerfeld coefficient $\gamma = 9.5$ mJ/mole K$^2$. No observable features that could be correlated to the anomalies in the $\chi(T)$ near 40 and 85 K could be detected. Measurements of the time dependence of the temperature of the sample stage of the calorimeter near 40 and 85 K didn't show any anomalies that could be due to first order phase transitions.

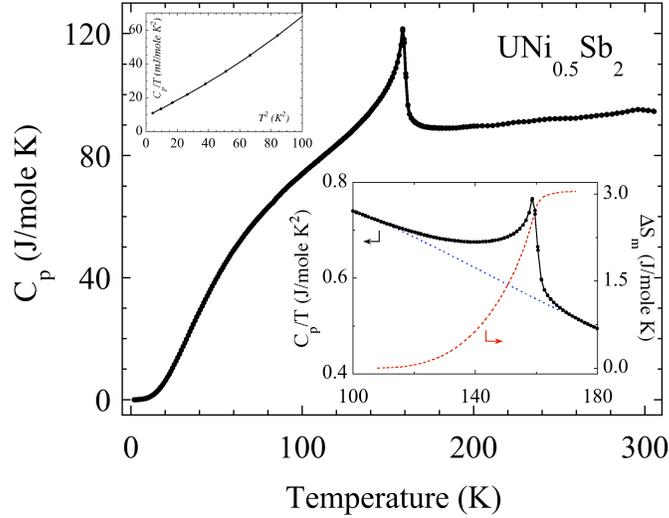

Figure 4 (color online) – $C_p(T)$ data for $UNi_{0.5}Sb_2$ between 1.8 and 300 K. The large peak centered at 161 K is due to the onset of AFM order. The lower inset shows how the magnetic entropy was extracted, yielding $\Delta S_{mag} \approx 3.0$ J/mole K, which corresponds to 52% of $R\ln 2$. The upper inset shows a plot of $C/T$ vs $T^2$ at low temperatures, yielding $\gamma \approx 9.5$ mJ/mole $K^2$.

The thermal conductivity $\kappa(T)$ for $\Delta T$ across the *ab*-plane is shown in Fig. 5. A shallow minimum near 150 K correlates somewhat in temperature with the onset of AFM, suggesting that phonon and electron scattering due to spin fluctuations are being suppressed near $T_N$. Whereas no distinct features were observed near 85 K, a small 3 K hysteresis between the cooling and warming cycle was observed near 40 K.

The temperature dependence of the thermopower, with the temperature gradient across the *ab*-plane, is shown in Fig. 6. The value of $S(T)$ drops sharply at $T_N$, and two hysteretic anomalies near 40 and 85 K correlate well in temperature with the features in the $\chi(T)$ data. The $S(T)$ data for $T > T_N$ fit well to the phenomenological expression $S(T) = AT/(B^2+T^2)$,[21] yielding $A = 13.4$ mV and $B = 247.8$ K. Since $A$ and $B$ are functions of the Fermi energy ($E_F$) and band width ($\Gamma$), it is not surprising that a sharp departure from the model takes place below $T_N$, due to the formation of a magnetic superzone and the concomitant change in Fermi surface topology. The thermopower is also quite sensitive to the two lower temperature features. Similarly to the $\chi(T)$ data, the two low temperature features show strong thermal hysteresis.

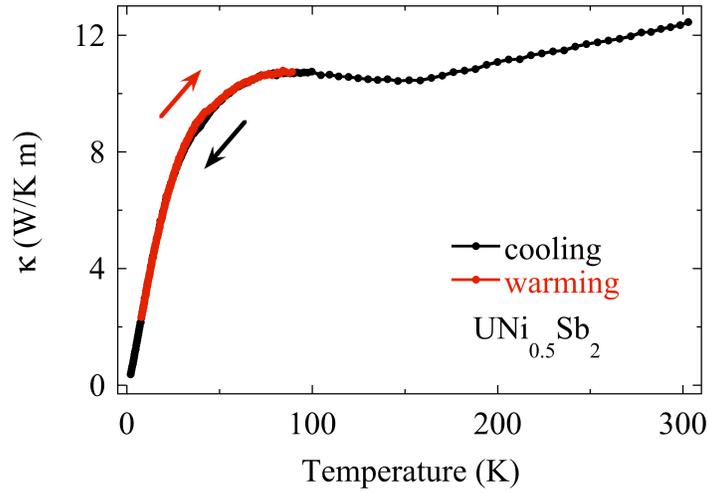

Figure 5 (color online) - Thermal conductivity for $UNi_{0.5}Sb_2$ between 1.8 and 300 K for $\Delta T$ across the *ab*-plane.

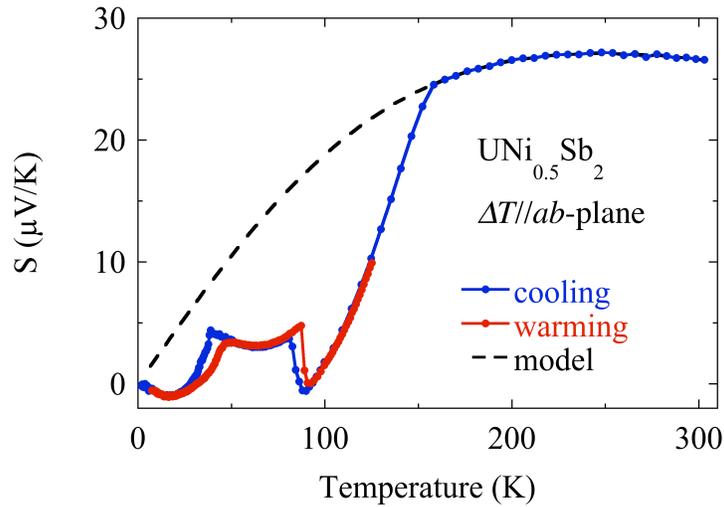

Figure 6 (color online) – Thermopower vs temperature for $UNi_{0.5}Sb_2$ between 1.8 and 300 K. The dashed line is a fit to the one-band model.

The temperature dependence of the linear thermal expansion $\Delta L/L$ along the *ab*-plane and the *c*-axis are shown in Fig. 7. There is a noticeable change in volume at $T_N$, and the two hysteretic features centered near 40 and 85 K correlate well in temperature with the features in magnetic susceptibility and thermopower. The maximum values of the changes in $\Delta L/L$ at 40 and 85 K, corresponding to the maximum values of the anomalies in $\Delta a/a$ and $\Delta c/c$, are $\approx 0.3 \times 10^{-4}$ and $1.0 \times 10^{-4}$, respectively. These values are very small, and they are far below the resolution limit of typical diffraction studies.

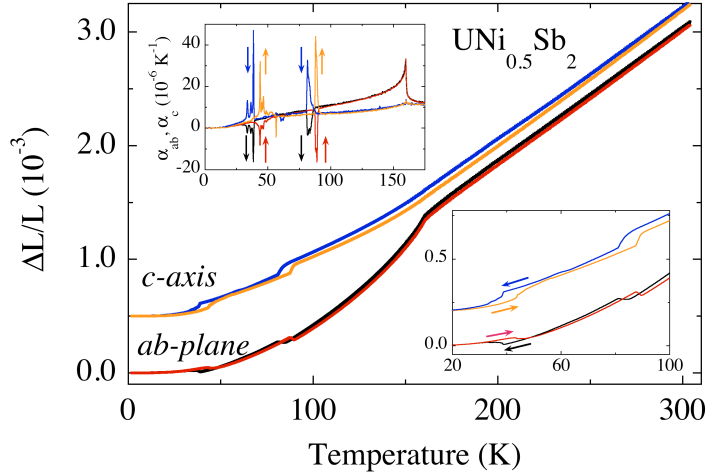

Figure 7 (color online) – Thermal expansion parallel and perpendicular to the $c$-axis in $UNi_{0.5}Sb_2$. The onset of AFM at 161 K, and the 2 lower temperature features can be clearly identified. The $\Delta L/L$ data for the $c$-axis is offset by $0.5 \times 10^{-3}$ (main panel) and $0.2 \times 10^{-3}$ (inset) for clarity. The lower inset details the hysteretic nature of the two features near 40 and 85 K upon cooling and warming. The upper inset shows the temperature dependence of the linear coefficients of thermal expansion $\alpha_{ab}$ and $\alpha_c$.

### 3.5- Electrical resistivity under pressure

The curves of normalized electrical resistivity $\rho_{ab}/\rho_{ab,\,300K}$ and $\rho_c/\rho_{c,\,300K}$ vs $T$ for various pressures are shown in Fig. 8. The $\rho(T)$ curves for $P > 0$ are offset for clarity. The $\rho_{ab}(T)$ data for $P = 0$ show several distinct features; 1) the temperature coefficient of the electrical resistivity changes from negative in the paramagnetic phase to positive below $T_N$, reflecting the loss of spin-disorder scattering across the $ab$-plane in the AFM phase; 2) the two small features centered near 40 and 85 K are both hysteretic. These two features remain unaffected by a 9 T magnetic field ($H//c$-axis; data not shown). The effect of pressure on the onset of AFM order is to raise the value of $T_N$ at the rate of $\approx 0.76$ K/kbar, as shown in the inset of Fig. 8a. In addition, pressure first lowers the onset temperature of the two low temperatures features upon cooling, while broadening the width of the hysteretic regions. The $\rho_{ab}$ vs $T$ data show that the 2 low temperature features start to overlap for $P \approx 14.8$ kbar, and they can no longer be distinguished for $P \approx 20.6$ kbar. The $\rho_c/\rho_{c,\,300K}$ vs $T$ data of Fig. 8b also show a distinct feature at $T_N$, which is also shifted to higher temperatures under pressure. The value of $\rho_c$ is nearly independent of the temperature in the paramagnetic phase. However, it shows a sharp upward discontinuity at $T_N$, consistent with the formation of a magnetic superzone energy gap, due to the alternate orientation of the U ions in neighboring *ab*-planes below $T_N$. Similarly to $\rho_{ab}$, the 2 low temperatures feature in $\rho_c$ are hysteretic upon the cooling and warming cycles. The effect of pressure is also to broaden them, shift them to lower temperatures, merge and eventually suppress them, to the point where they can hardly be distinguished for $P \approx$ 21.4 kbar.

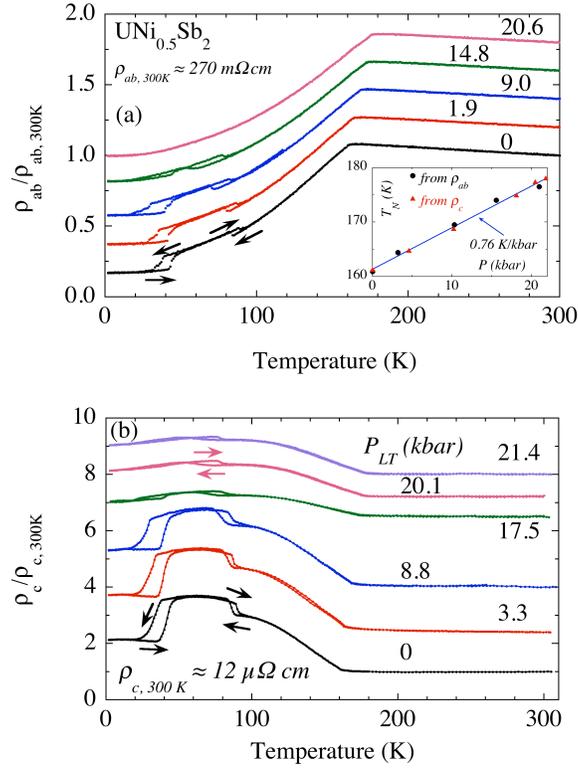

Figure 8 (color online) – Normalized electrical resistivity $\rho/\rho_{300K}$ versus temperature for UNi$_{0.5}$Sb$_2$ in pressures up to 22 kbar for (a) *I*//*ab*-plane; and (b) *I*//*c*. The curves for $P > 0$ are offset for clarity. The discontinuity in $\rho_{ab}$ and $\rho_c$ near 161 K is due to the onset of AFM order, and it is shifted upwards with $P$, while the 2 hysteretic features centered near 40 and 85 K are suppressed. The inset of the (a) pane shows the $P$ dependence of $T_N$. The pressures indicated in both panes are the values for $T < 90$ K. The $P$ values between ambient temperature and $T > 90$ K were estimated by linear interpolation (see text).

## 4- Discussion and conclusions

The structural refinement using the neutron diffraction data showed that the space group that better describes the Sb flux-grown $UNi_{0.5}Sb_2$ single crystals is $P4_2/nmc$. This structure is quite similar to the $HfCuSi_2$-type ($P4/nmm$); the $P4_2/nmc$ can be obtained from the $P4/nmm$ structure by doubling along $c$, and translating the origin of the unit-cell. There are two Ni sites in the $P4_2/nmc$ structure, and in the case of $UNi_{0.5}Sb_2$, one is nearly fully occupied and the other is nearly empty. The neutron diffraction data at 25 K are consistent with this same structure, and it shows additional reflections and intensities, which are consistent with an AFM ground state with ordered U-moment values $\approx 1.85 \pm 0.12$ $\mu_B$. It should be noted that 25 K, the temperature of the magnetic structural neutron diffraction measurements, is much below $T_N$, and somewhat below the $\approx 40$ K and $\approx 85$ K temperature where the anomalies in the bulk properties were observed. Although the thermal expansion data suggests the occurrence of two minor structural distortions near $\approx 40$ K and 85 K, these distortions are too small to be detected in typical diffraction studies.

The two hysteretic features in $\chi(T)$, $S(T)$, $\rho(T)$, and $\Delta L/L$ near 40 and 85 K are consistent with intrinsic behavior, and they involve changes in Fermi surface topology. It is tempting to consider subtle changes in occupancy at the Ni sites, and/or some sort of spin or crystallographic realignment as the origin of these low-$T$ features. In light of the anomalies of the thermal expansion near 40 and 85 K, magnetoelastic interactions driven by these changes should be considered as well. However, these rearrangements are not drastic enough to be picked up by the specific heat measurements, or to leave an imprint in the neutron diffraction data at 25 K. It is clear from the $\rho(T, P)$ data, that pressure suppresses the transformations at 40 and 85 K, stabilizing the high-$T$ phase. The estimated change in magnetic entropy due to the onset of AFM order is $\approx 3.0$ J/mole K, corresponding to $0.52R\ln2$, which is consistent with the proposed long-range magnetic order. Thermopower measurements are particularly sensitive to changes in Fermi surface topology, and in the $UNi_{0.5}Sb_2$ case, they suggest that quite dramatic changes take place at $T_N$ and at the lower temperature features. It is plausible that additional neutron diffraction studies in the temperature range of the two low-$T$ features might shed some light into their origin.

The initial hydrostatic pressure derivative $dT_N/dP$ can be estimated using the Ehrenfest relation for second order phase transitions [Ref. 22]

$$\frac{dT_N}{dP} = V_m T_N \frac{\Delta\beta}{\Delta C}$$

where $V_m$ is the molar volume, and $\Delta\beta$ and $\Delta C$ are the changes in the volumetric thermal expansion coefficient and heat capacity at $T_N$, respectively. Using the experimental values from our measurements, $V_m = 5.05 \times 10^{-5}$ m$^3$, $T_N = 161$ K, $\Delta\beta = 2\Delta\alpha_{ab} + \Delta\alpha_c = 36.7 \times 10^{-6}$ K$^{-1}$, and $\Delta C = 32.3$ J/mole K, we obtain $dT_N/dP = 0.92$ K/kbar. This value is about 20% higher than the value determined from the $\rho(T)$ measurements under pressure; considering the experimental uncertainties in the determination $\Delta\beta$ and $\Delta C$, this can be regarded as being in good agreement. Using a similar Ehrenfest relation, with $\Delta\alpha_{ab}$ (17.7 x 10$^{-6}$ K$^{-1}$) or $\Delta\alpha_c$ (1.36 x 10$^{-6}$ K$^{-1}$) substituting for $\Delta\beta$, the initial uniaxial pressure derivatives $dT_N/dP_{ab}$ and $dT_N/dP_c$ can be estimated, yielding 0.445 and 0.034 K/kbar, respectively. The uniaxial pressure response is very anisotropic, with the in-plane value being one order of magnitude higher than along the $c$-axis.

It is quite puzzling that the hysteretic anomalies below $T_N$ in the samples from two groups are centered near 40 and 85 K, while the samples from another group show one anomaly centered near 60 K. Also, the $C_p(T)$ data in the latter showed a small but clear hysteretic feature near 60 K, while the samples with two low-$T$ anomalies don't show any distinguishable corresponding features in $C_p(T)$. Although the origin of these low-$T$ anomalies is still elusive, it is quite clear from the $\rho(T, P)$ data that the effect of pressure is to stabilize the high-$T$ phase.

The behavior of $\rho_{ab}(T)$ and $\rho_c(T)$ near $T_N$ is consistent with the proposed type of AFM order, and the two anomalies at the lower temperatures are consistent with the changes in Fermi surface suggested by the $S(T)$ measurements. The effect of the hydrostatic pressure on $T_N$ is to raise it at the rate of $\approx 0.76$ K/kbar. The $\chi(T)$ curves for $H//ab$-plane and $H//c$ yield Curie-Weiss temperatures $\Theta \approx 100$ and -240 K, respectively, consistent with the AFM arrangement of Fig. 2. The effective magnetic moments inferred from the fits of the $\chi^{-1}$ vs $T$ data at high temperatures to a Curie-Weiss behavior were $\approx$ 3.15 $\mu_B$ ($H//c$) and $\approx$ 3.25 $\mu_B$ ($H//ab$-plane), both reduced from the 3.58 and 3.62 $\mu_B$ values expected for U$^{4+}$ and U$^{3+}$, respectively. In contrast, the effective moment for the arc melted polycrystalline specimen of Ref. [1], with nominal composition UNiSb$_2$, was 2.46 $\mu_B$, suggesting a higher hybridization level in the latter, consistent with a higher occupancy of the Ni sites. If this argument is valid, and considering that 1) $T_N \approx 175$ K for the UNiSb$_2$ polycrystalline sample, 2) $T_N \approx 161$ K for the UNi$_{0.5}$Sb$_2$ single crystals, and 3) that pressure promotes hybridization while raising $T_N$ in the UNi$_{0.5}$Sb$_2$ at the rate of $\approx 0.76$ K/kbar, we have the unique situation in which the increased hybridization is concurrent with an increase in the exchange interaction that supports the AFM order. Studies in higher pressures should shed some light in this issue.

*Acknowledgments* - The support from NSF Grant No. DMR-0805335 for the work at SDSU, NSF Grant No. DMR-0804032 for the work at NMSU, USDOE Contract DE-AC02-07CH11358 for work at Ames Laboratory and ISU, and USDOE Contract DE-AC52-06NA25396 for work at the Manuel Lujan, Jr. Neutron Scattering Center at Los Alamos National Laboratory are gratefully acknowledged.